\documentstyle[epsf,rotate]{apjmn}
 at 10truept
\title
     [Eigenvalue Problems]
{\vglue-1.7truecm
\centerline{\footnotesize {\it ApJ}, in press.
Available from {\it h t t p://www.mpa-garching.mpg.de/$\tilde{~}$max/karhunen.html} (faster from Europe)}
\vglue-0.3truecm\centerline{\footnotesize
and from {\it h t t p://www.sns.ias.edu/$\tilde{~}$max/karhunen.html} (faster from the US).}
\vglue 1.2truecm
\noindent
	Karhunen-Lo\`eve Eigenvalue Problems in Cosmology:\\
	How should we Tackle Large Data Sets?
}	
\author
     [M. Tegmark, A. N. Taylor, \& A. F. Heavens]
     {Max Tegmark$^{1,2,4}$, Andy N. Taylor$^3$, \& Alan F. Heavens$^3$ \\
     $^1$Institute for Advanced Study, Olden Lane, Princeton, NJ 08540\\
     $^2$Max-Planck-Institut f\"{u}r Astrophysik, 
     Karl-Schwarzschild-Str. 1, D-85740 Garching\\
     $^3$Institute for Astronomy, 
     University of Edinburgh,
     Royal Observatory,
     Blackford Hill, 
     Edinburgh, 
     U.K.\\
     $^4$Hubble Fellow\\
     max@ias.edu, ant@roe.ac.uk, afh@roe.ac.uk}
\def\bib{\parskip=0pt\par\noindent\hangindent\parindent
    \parskip =2ex plus .5ex minus .1ex}
\newcommand{\be}{\begin{equation}}
\newcommand{\ee}{\end{equation}}
\newcommand{\ba}{\begin{eqnarray}}
\newcommand{\ea}{\end{eqnarray}}

\newcommand{\etal}{\frenchspacing \em et al.}
\newcommand{\ie}{{\frenchspacing \em i.e.}}
\newcommand{\eg}{{\frenchspacing \em e.g.}}
\newcommand{\etc}{{\frenchspacing \em etc.}}
\newcommand{\rms}{{\frenchspacing \em r.m.s.}}

\newcommand{\x}{{\bmath x}}
\newcommand{\y}{{\bmath y}}
\newcommand{\z}{{\bmath z}}
\newcommand{\A}{{\bmath A}}
\newcommand{\B}{{\bmath B}}
\newcommand{\C}{{\bmath C}}
\newcommand{\D}{{\bmath D}}
\newcommand{\F}{{\bmath F}}
\newcommand{\I}{{\bmath I}}
\newcommand{\bP}{{\bmath P}}

\newcommand{\Sig}{{\bmath S}}
\newcommand{\T}{{\bmath T}}
\newcommand{\M}{{\bmath M}}
\newcommand{\N}{{\bmath N}}
\newcommand{\U}{{\bmath U}}
\newcommand{\V}{{\bmath V}}
\newcommand{\bL}{{\bmath L}}

\newcommand{\Tr}{{\rm Tr}}

\font\bfmath=cmmib10
\def\vth{
\hbox{\bfmath\char'002}}  	
\def\vmu{
\hbox{\bfmath\char'026}}	
\def\bLambda{
\hbox{\bfmath\char'0003}}	
\def\bPhi{
\hbox{\bfmath\char'010}}	
\def\beq#1{\begin{equation}\label{#1}}
\def\eeq{\end{equation}}
\def\beqa#1{\begin{eqnarray}\label{#1}}
\def\eeqa{\end{eqnarray}}
\def\eq#1{equation~(\ref{#1})}
\def\Eq#1{Equation~(\ref{#1})}

\def\bfig{\begin{figure}[h] \centerline{\hbox{}}\vfill}
\def\efig{\end{figure}\vfill\newpage}

\def\fig#1{Figure~\ref{#1}}
\def\Fig#1{Figure~\ref{#1}}

\def\mK{{\rm \mu K}}

\def\Mpc{{\rm Mpc}}

\def\vthml{\vth_{\hbox{ML}}}
\def\expec#1{\langle#1\rangle}
\def\bigexpec#1{\left\langle#1\right\rangle}
\def\diag{{\rm diag}}
\def\Ell{{\cal L}}
\def\l{\ell}
\def\lmax{\l_{\hbox{\footnotesize max}}}
\def\beamsig{\theta_b}
\def\FWHM{{\rm FWHM}}
\def\crr{\cr\noalign{\vskip 4pt}}
\def\alm{a_{\l m}}
\def\Cl{C_\l}

\def\vb{{\bmath b}}
\def\nh{\hat{\bf n}}
\def\Fnew{{\tilde{\F}}}
\def\ith{i^{\rm th}}
\def\kth{k^{\rm th}}

\def\spose#1{\hbox to 0pt{#1\hss}}
\def\simlt{\mathrel{\spose{\lower 3pt\hbox{$\mathchar"218$}}
     \raise 2.0pt\hbox{$\mathchar"13C$}}}
\def\simgt{\mathrel{\spose{\lower 3pt\hbox{$\mathchar"218$}}
     \raise 2.0pt\hbox{$\mathchar"13E$}}}
\def\simpropto{\mathrel{\spose{\lower 3pt\hbox{$\mathchar"218$}}
     \raise 2.0pt\hbox{$\propto$}}}

\begin{document}
 
\maketitle

\begin{abstract}

Since cosmology is no longer ``the data-starved science", 
the problem of how to best analyze large data sets has recently 
received considerable attention, and 
Karhunen-Lo\`eve eigenvalue methods have been applied
to both galaxy redshift surveys and 
Cosmic Microwave Background (CMB) maps. 
We present a comprehensive discussion of methods for 
estimating cosmological parameters from large data sets, which 
includes the previously published techniques as special cases.
We show that both the problem of estimating several parameters
jointly and the problem of not knowing the parameters 
a priori can be readily solved by adding an extra singular 
value decomposition step. 

It has recently been argued that the information content in a
sky map from a next generation CMB satellite 
is sufficient to measure key cosmological parameters 
($h$, $\Omega$, $\Lambda$, {\etc}) to an accuracy of a 
few percent or better --- in principle. 
In practice, the data set is so large that both a brute force 
likelihood analysis and a 
direct expansion in signal-to-noise eigenmodes will be
computationally unfeasible.
We argue that it is likely that 
a Karhunen-Lo\`eve approach can nonetheless measure the 
parameters with close to maximal accuracy,
if preceded by an appropriate form of quadratic ``pre-compression''. 

We also discuss practical issues regarding parameter estimation from
present and future galaxy redshift surveys, and illustrate this
with a generalized eigenmode analysis of the IRAS 1.2 Jy 
survey optimized for measuring $\beta\equiv\Omega^{0.6}/b$
using redshift space distortions.

\end{abstract}

\begin{keywords} 
Cosmology: theory, large scale structure of Universe, cosmic 
microwave background -- Astronomical methods: data analysis, statistical 
\end{keywords}

\section{INTRODUCTION}
\label{IntroSec}

The problem of analysis of large data sets is one that, until
recently, has not been a major concern of cosmologists.
Indeed,
in some areas no data existed to be analyzed. In the last 
few years, this situation has rapidly changed. A highlight in this
transition has been the discovery of fluctuations
in the cosmic microwave background (CMB) by the 
Cosmic Background Explorer (COBE) satellite
(Smoot {\etal} 1992). In its
short lifetime, COBE produced such a large data set that
a number of sophisticated data-analysis methods 
were developed specifically to tackle it.
In addition, the advent of large 
galaxy redshift surveys has created a field where the data sets 
increase by an order of magnitude in size in each generation.  
For instance, the surveys where the object selection was based on 
the Infrared Astronomy Satellite (IRAS) contain 
several thousand galaxies: $\sim 2,000$ (QDOT; Lawrence {\etal} 1996) 
$\sim 5,000$ (Berkeley 
1.2Jy; Fisher {\etal} 1995) and $\sim 15,000$  (PSC-z; 
Saunders {\etal} 1996).   
The proposed next-generation surveys will 
have much larger numbers of objects --- around 250,000 in the Anglo-Australian
Telescope 2 degree Field galaxy redshift survey (Taylor 1995) and 
$\sim 10^6$ for the
Sloan Digital Sky Survey (Gunn \& Weinberg 1995).  
Similarly plate measuring machines, such as the APM at Cambridge and 
SuperCOSMOS at the Royal Observatory, Edinburgh,
can produce very large catalogues of objects, and numerical simulations of 
galaxy clustering are even now capable of producing so much data that the 
analysis and storage of the information is in itself a challenge. 

A standard technique for estimating parameters from data is the 
brute force maximum likelihood method, 
which illustrates why people have been driven towards
developing more sophisticated methods. 
For $n$ data items ({\eg}, pixels in a CMB map, or Fourier 
amplitudes from a transformed galaxy distribution), 
the maximum likelihood
method requires inversion of an $n\times n$ matrix for each set of parameter
values considered --- and this is for the simplest possible case where
the probability distribution is Gaussian.
Since a next-generation CMB satellite might produce a high resolution sky map
with $\sim 10^7$ pixels, and the CPU time required for an inversion scales
as $n^3$, a brute force likelihood analysis of this type of data set
will hardly be feasible in the near future.

Fortunately, it is often possible to greatly accelerate a likelihood 
analysis by first performing some appropriate form of data compression,
by which the data set is substantially reduced in size while nonetheless
retaining virtually all the relevant cosmological information. 
In this spirit, a large number of data-compression methods have been 
applied in the analysis of both CMB maps 
({\eg} Seljak \& Bertschinger 1993; G\'orski 1994)
and galaxy redshift surveys
({\eg} Davis \& Peebles 1983; Feldman {\etal} 1994; Heavens \& Taylor 1995).
A powerful prescription for how to do this optimally is the 
Karhunen-Lo\`eve eigenvalue method (Karhunen 1947), 
which has recently been applied to both 
CMB maps (Bond 1994; Bunn 1995; Bunn \& Sugiyama 1995) 
and redshift surveys (Vogeley 1995; Vogeley \& Szalay 1996). 
The goal of this paper is to review the more general framework in
which these treatments belong, and to present some important generalizations 
that will facilitate the analysis of the next generation of 
cosmological data sets.

The rest of this paper is organized as follows.
In Section~\ref{FisherSec}, we review some useful information-theoretical
results that tell us how well parameters can be estimated, and
how to determine whether a given analysis method is good or bad.
In Section~\ref{CompressionSec}, we review the 
Karhunen-Lo\`eve data compression method and present 
some useful generalizations.
In Section~\ref{ApplicationsSec} we illustrate the theory with various
cosmology applications, including the special case of 
the signal-to-noise eigenmode method.
In Section~\ref{PreCompressionSec}
we discuss limitations of method and possible ways of 
extending it to make the analysis feasible for huge data sets such as a  
$10^7$ pixel future CMB map. 
Finally, our results are summarized and discussed in 
Section~\ref{ConclusionsSec}.

\section{HOW WELL CAN YOU DO WITHOUT DATA COMPRESSION?}
\label{FisherSec}

How accurately can we estimate model parameters from a given data set?
This question was basically answered
60 years ago (Fisher 1935), and we will now summarize the results, which are 
both simple and useful.

\subsection{The classical results}

Suppose for definiteness that our data set consists on $n$ real numbers
$x_1,\>x_2,...,x_n$, which we arrange in an $n$-dimensional 
vector $\x$. These numbers could for instance denote the measured 
temperatures in the $n$ pixels of a CMB sky map, 
the counts-in-cells
of a galaxy redshift survey, the first $n$ coefficients
of a Fourier-Bessel expansion of an observed galaxy 
density field, 
or the number of gamma-ray bursts observed in $n$ different flux bins.
Before collecting the data, we think of $\x$ as a random variable
with some probability distribution $L(\x;\vth)$, which 
depends in some known way on a vector of model parameters
\beq{ThetaDefEq}
\vth= (\theta_1, \theta_2, ..., \theta_m).
\eeq
Such model parameters might for instance be 
the spectral index of density fluctuations, the Hubble constant $h$, 
the cosmic density parameter $\Omega$ or the mean redshift of gamma-ray bursts.
We will let $\vth_0$ denote the true parameter values and 
let $\vth$ refer to our estimate of $\vth$. 
Since $\vth$ is some function of the data vector $\x$, it too is a random variable.
For it to be a good estimate, we would of course like it to be unbiased,
{\ie}, 
\beq{BiasEq}
\expec{\vth} = \vth_0,
\eeq
and give as small error bars as possible, {\ie}, minimize the
standard deviations
\beq{SdevEq}
\Delta\theta_i\equiv\left(\bigexpec{\theta_i^2}-\expec{\theta_i}^2\right)^{1/2}
\eeq
In statistics jargon, we want the BUE $\theta_i$, which stands for the  
``Best Unbiased Estimator".

A key quantity in this context is the so-called 
{\it Fisher information matrix}, defined as 
\beq{FisherDefEq}
\F_{ij} \equiv \bigexpec{{\partial^2\Ell\over\partial\theta_i\partial\theta_j}},
\eeq
where $\Ell\equiv -\ln L$.
Another key quantity is the {\it maximum likelihood estimator}, or
{\it ML-estimator} for brevity, defined as the
parameter vector $\vthml$ that maximizes the likelihood function $L(\x;\vth)$.
(Above we thought of $L(\x;\vth)$ as a probability distribution, a function of $\x$.
However, when limiting ourselves to a given data set $\x$ and 
thinking of $L(\x;\vth)$ as a function of the parameters $\vth$, it 
is customarily called the {\it likelihood function}.)

Using this notation, a number of powerful theorems have been proven
(see {\eg} Kenney \& Keeping 1951 and Kendall \& Stuart 1969 for details):
\begin{enumerate}

\item For any unbiased estimator, $\Delta\theta_i \geq 1/\sqrt{\F_{ii}}$.

\item If there is a BUE $\vth$, then it is the ML-estimator 
or a function thereof.

\item The ML-estimator is asymptotically BUE.

\end{enumerate}
The first of these theorems, 
known as the Cram\'er-Rao inequality, thus places 
a firm lower limit on the 
error bars that one can attain, regardless of which method one is
using to estimate the parameters from the data. 
This is the minimum error bar attainable on $\theta_i$ if all the
other parameters are known. If the other parameters are estimated
from the data as well, the minimum standard deviation 
rises to $\Delta\theta_i \geq (\F^{-1})_{ii}^{1/2}$.

The second theorem 
shows that maximum-likelihood (ML) estimates have quite a special status:
if there is a best method, then the ML-method is the one.
Finally, the third result basically tells us that in the limit of a 
very large data set, the ML-estimate for all practical purposes is 
the best estimate, the one that for which the 
Cram\'er-Rao inequality becomes an equality.
It is these nice properties that have made 
ML-estimators so popular. 
 
\subsection{The Fisher information matrix}

Although the proof of the Cram\'er-Rao inequality is 
rather lengthy,
it is quite easy to acquire some intuition
for why the Fisher information matrix has the form that it does.  
This is the purpose of the present section.

Let us Taylor expand $\Ell$ around the ML-estimate $\vth$.
By definition, all the first derivatives $\partial\Ell/\partial\theta_i$ 
will vanish at the ML-point, since the likelihood function has its maximum there,
so the 
local behavior will be dominated by the quadratic terms. 
Since $L=\exp[-\Ell]$, 
we thus see that the likelihood 
function will be approximately 
Gaussian near the ML-point.
If the error bars are quite small, $L$
usually drops sharply before third order terms 
have become important, so that this Gaussian is a good
approximation to $L$ everywhere.
Interpreting $L$ as a Bayesian probability distribution
in parameter space, the covariance matrix 
$\T$ is thus given simply by the second derivatives
at the ML-point, as the inverse of the Hessian matrix:
\beq{HdefEq} 
(\T^{-1})_{ij}\equiv{\partial^2\Ell\over\partial\theta_i\partial\theta_j}.
\eeq
Note that the Fisher information matrix $\F$ is simply the expectation
value of this quantity at the point $\vth=\vth_0$
(which coincides with the ML-point on average if the 
ML-estimate is unbiased).
It is basically a measure of how fast (on average) the likelihood function
falls off around the ML-point, {\ie}, a measure of the width
and shape of the peak. From this discussion, 
it is also clear that we can use its inverse, $\F^{-1}$, as an estimate  
of the covariance matrix 
\beq{ParCovEq}
\T\equiv \expec{\vth\vth^t} - \expec{\vth}\expec{\vth}^t
\eeq
of our parameter estimates when we use the ML-method.

\subsection{The Gaussian case}

Let us now explicitly compute the Fisher information matrix for the case 
when the probability distribution is Gaussian, {\ie}, where 
(dropping an irrelevant additive constant $n\ln[2\pi]$)
\beq{GaussianEq}
2\Ell = \ln\det \C + (\x-\vmu) \C^{-1}(\x-\vmu)^t,
\eeq
where in general both the mean vector $\vmu$ and the covariance matrix
\beq{covdefEq}
\C = \expec{(\x-\vmu)(\x-\vmu)^t}
\eeq 
depend on the model parameters $\vth$.
Although vastly simpler than the most general situation, the Gaussian case 
is nonetheless general
enough to be applicable to a wide variety of problems in cosmology. 
Defining the data matrix
\beq{DdefEq} 
\D\equiv (\x-\vmu)(\x-\vmu)^t
\eeq
and using the well-known matrix identity 
$\ln\det \C = \Tr\ln \C$, 
we can re-write \eq{GaussianEq} as
\beq{GaussianEq2}
2\Ell = \Tr\left[\ln \C + \C^{-1}\D\right].
\eeq
We will use the standard comma notation for derivatives, where for instance
\beq{CommaDefEq}
\C,_i \equiv {\partial\over\partial\theta_i}\C.
\eeq
Since $\C$ is a symmetric matrix for all values of the 
parameters, it is easy to see that
all the derivatives $\C,_i$, $\C,_{ij}$, {\etc}, will also 
be symmetric matrices. Using the 
matrix identities $(\C^{-1}),_i = -\C^{-1}\C,_i \C^{-1}$ and 
$(\ln \C),_i = \C^{-1} \C,_i$, we thus obtain
\beq{LiEq}
2\Ell,_i = \Tr\left[\C^{-1} \C,_i -
		 \C^{-1}\C,_i \C^{-1}\D + \C^{-1}\D,_i\right].
\eeq
When evaluating $\C$ and $\vmu$ at the true parameter values, 
we have $\expec{\x}=\vmu$ and $\expec{\x\x^t}=\C+\vmu\vmu^t$, which 
gives
\beq{DexpecEq}
\cases{
\expec{\D}	&$=\C$,\crr
\expec{\D,_i}	&$=0$,\crr
\expec{\D,_{ij}}	&$=\vmu,_i\vmu,_j^t + \vmu,_j\vmu,_i^t$.\crr
}
\eeq
Using this and \eq{LiEq}, we obtain $\expec{\Ell,_i} = 0$.
In other words, the ML-estimate
is correct on average in the sense that the average slope 
of the likelihood function is zero at the point 
corresponding to the true parameter values.
Applying the chain rule to \eq{LiEq}, we obtain
\beqa{LijEq}
\nonumber
2\Ell,_{ij} = \Tr[
&-&\C^{-1}\C,_i\C^{-1}\C,_j 
 + \C^{-1} \C,_{ij}\\
\nonumber
&+& \C^{-1}(\C,_i \C^{-1} \C,_j + \C,_j \C^{-1} \C,_i)\C^{-1} \D\\ 
\nonumber
&-& \C^{-1}(\C,_i \C^{-1} \D_{,j}+\C,_j \C^{-1} \D_{,i})\\
&-& \C^{-1}\C,_{ij} \C^{-1} \D 
 + \C^{-1} \D,_{ij}
].
\eeqa
Substituting this and \eq{DexpecEq} into \eq{FisherDefEq} and using the 
trace identity
$\Tr [\A\B] = \Tr [\B\A]$, many terms drop out and 
the Fisher information matrix reduces to simply
\beq{GaussFisherEq}
\F_{ij} = \expec{\Ell,_{ij}} 
= {1\over 2}\Tr[\A_i \A_j + \C^{-1}\M_{ij}],
\eeq
where we have defined the matrices $\A_i\equiv \C^{-1} \C,_i=(\ln \C),_i$
and $\M_{ij}\equiv \expec{\D,_{ij}}= \vmu,_i\vmu,_j^t + \vmu,_j\vmu,_i^t$.
This old and well-known result is also derived by Bunn (1995) and Vogeley
\& Szalay (1996).

\subsection{An example: parameter estimation with a future CMB experiment}

Let us illustrate the above results with a simple example
where $\vmu=0$ and $\C$ is diagonal.
Suppose a next generation CMB satellite generates a high-resolution
all-sky map of the CMB fluctuations. Let us for the moment ignore the 
complication of the galactic plane, and expand the temperature
distribution in spherical harmonics $\alm$. 
We define our data vector $\x$ as 
the set of these coefficients from $\l=2$ up to 
some cutoff $\lmax$, {\ie},
$n = (\lmax+1)^2 - 4$ and 
$x_i = a_i$, 
where we have combined $\l$ and $m$ into a single index 
$i=1,2,3...$ as 
$i=\l^2+\l+m+1$.
With the standard assumption that 
the CMB fluctuations are isotropic, we have
\beq{almEq} 
\cases{
\vmu	&$=0$,\crr
\C_{ij}	&$=\delta_{ij}\left[\Cl + {4\pi\sigma^2\over N} e^{\beamsig^2\l(\l+1)}\right]$.
}
\eeq
Here $\Cl$ denotes the angular power spectrum of the CMB, 
and the second term incorporates the effects of instrumental noise and 
beam smearing (Knox 1995, Tegmark \& Efstathiou 1996).
$N$ is the number of pixels in the sky map, $\sigma$ is the 
{\rms} pixel noise, 
\beq{FWHMeq}
\beamsig\equiv\FWHM/\sqrt{8\ln 2} \approx 0.425\>\FWHM,
\eeq
and $\FWHM$ is the full-width-half-max beam width.
Assuming that the CMB and noise fluctuations are Gaussian, 
\eq{GaussFisherEq}
gives the Fisher information matrix 
\beq{JungmanEq}
\F_{ij} =
\sum_{\l=2}^{\lmax} \left({2\l+1\over 2}\right)
\left[C_\l + {4\pi\sigma^2\over N} e^{\beamsig^2\l(\l+1)}\right]^{-2}
\left({\partial\Cl\over\partial\theta_i}\right)
\left({\partial\Cl\over\partial\theta_j}\right).
\eeq

\begin{figure}
\centerline{\epsfysize=9cm\epsfbox{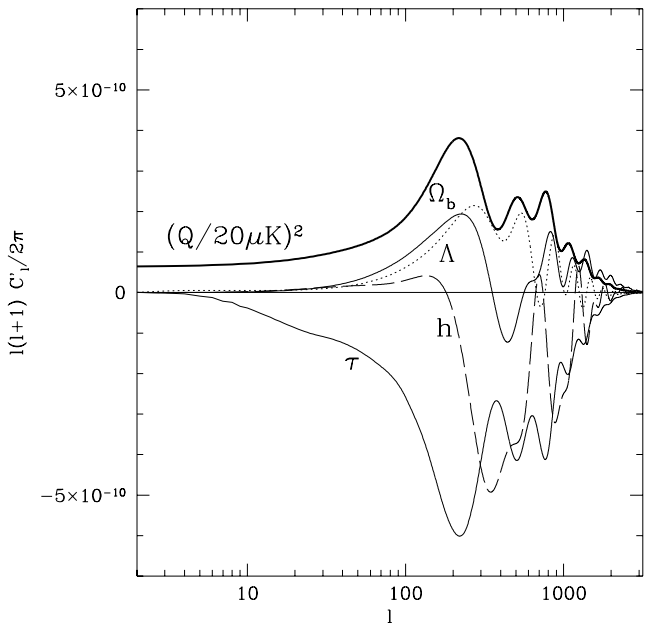}}
\caption{\label{DerivativesFig}
The derivatives of the CDM power spectrum with respect to various parameters.
}
\end{figure}

\noindent
This handy expression, also derived by 
Jungman {\etal} (1996a), tells us that the
crucial functions which determine the attainable accuracy are the
{\em derivatives} of the power spectrum with respect to the various parameters.
Examples of such derivatives are shown in \fig{DerivativesFig}.
For instance, the derivative with respect to the
reionization optical depth, 
$\partial\Cl/\partial\tau$, is shaped as $-\Cl$ for 
$\l\gg 10$, since earlier reionization suppresses all these multipoles
by the same factor $e^{-2\tau}$. 
The derivative with respect to the quadrupole normalization,
$\partial\Cl/\partial Q$, of course 
has the same shape as the power spectrum itself.

\Eq{JungmanEq} has a simple geometric interpretation.
We can think of the $m$ functions $\partial\Cl/\partial\theta_i$ as 
vectors in a Hilbert space of dimension $(\lmax-1)$, and think of 
the Fisher matrix component $\F_{ij}$ as simply the dot product of the vectors
$\partial\Cl/\partial\theta_i$ and $\partial\Cl/\partial\theta_i$.
This dot product gives a rather small weight to low 
$\l$-values, essentially because there are fewer $m$-modes there 
and correspondingly larger cosmic variance.
In addition, the weights are seen to be exponentially
suppressed for large $\l$, where beam smearing causes the effect of 
pixel noise to explode. 
If the parameter dependence of $\Cl$ was such that all 
$n$ vectors $\partial\Cl/\partial\theta_i$ were orthogonal
under this dot product, then $\F$ and the parameter covariance matrix 
$\T=\F^{-1}$ would be diagonal, and the errors in the estimates of the
different parameters would be uncorrelated.
The more similarly shaped two curves in \fig{DerivativesFig} are,
the harder it will be to break the degeneracy between the 
corresponding two parameters. In the extreme case where two curves 
have identical shape (are proportional to each other), the Fisher 
matrix
becomes singular and the resulting error bars on the two parameters
become infinite. It is therefore interesting to diagonalize the matrix 
$\T$ (or equivalently, its inverse $\F$). 
The eigenvectors will tell us which $m$ parameter combinations
can be independently estimated from the CMB data, and the corresponding 
eigenvalues tell us how accurately this can be done. 

It has been pointed out (Bond {\etal} 1994) that there will 
be a considerable 
parameter degeneracy if data is only available up to around the first
Doppler peak. This is clearly illustrated in \fig{DerivativesFig}: 
most of the curves lack strong features in this range, so
some of them can be well approximated by linear combinations of the others.
If a CMB experiment has high enough resolution to measure
the power out to $l\sim 10^3$, however, this degeneracy is broken.
Jungman {\etal} (1996b) compute the Fisher matrix for 
a model with the eleven unknown parameters 
\beq{ParDefEq}
\vth = (\Omega,\Omega_b h^2,h,\Lambda,n_S,\alpha,n_T,T/S,\tau,Q,N_\nu);
\eeq
the density parameter, 
the baryon density,
the Hubble parameter,
the cosmological constant, 
the spectral index of scalar fluctuations,
the ``running" of this index (a measure of the deviation from 
power law behavior), 
the spectral index of tensor fluctuations,
the quadrupole tensor-to-scalar ratio, 
the optical depth due to reionization, 
and the number of light neutrino species,
respectively.
They find that even when estimating all these parameters simultaneously, 
the resulting error bars on some of the key parameters 
are of the order of a few percent or better
for experiments with a resolution good enough to probe the first
few Doppler peaks. 
Even the abundance of hot dark matter
can be constrained in this fashion (Dodelson {\etal} 1996).
An up-to-date discussion of which additional parameters
can be constrained when using large-scale-structure data as well
is given by Liddle {\etal} (1996).

In should be stressed that the formalism above only tells us what the attainable 
accuracy is if the truth is somewhere near the point in parameter
space at which we compute the power derivatives.
\Fig{DerivativesFig} corresponds to standard COBE-normalized CDM, {\ie}, to 
\beq{JungmanParEq}
\vth_0 = (1,0.015,0.5,0,1,0,0,0,0,20\mK,3).
\eeq
The worst scenario imaginable would probably be extremely early reionization, 
since $\tau\gg 1$ would eliminate almost all small-scale 
power and introduce a severe degeneracy by 
making all the power derivatives almost indistinguishable
for $\l\gg 10$, the region where they differ 
the most in \fig{DerivativesFig}.

Needless to say, accurate parameter estimation also requires that 
we can compute the theoretical 
power spectrum accurately. It has been argued 
(Hu {\etal} 1995) 
that this can be done to better than $1\%$, by accurately modeling various 
weak physical effects such as Helium recombination and polarization. 
We will return to other real-world issues, such as foreground contamination
and incomplete sky coverage, further on.

\section{OPTIMAL DATA COMPRESSION: KARHUNEN-LO\`EVE METHODS}
\label{CompressionSec}

Above we saw how small error bars one could obtain 
when using all the information in the data set.
We also saw that for large data sets, these minimal 
error bars could be approximately attained by making 
a likelihood analysis using the entire data set.
However, there are of course many situations where this approach 
is not feasible even in the simple Gaussian case.

\subsection{The need for data compression}

If we wish to estimate $m$ parameters from $n$ data points, 
this would involve inverting an $n\times n$ matrix at a dense grid
of points in the $m$-dimensional parameter space. 
The CPU time required to invert such a 
non-sparse matrix scales as $n^3$, and the number of grid points
grows exponentially with $m$, so there are obviously 
limits as to what can be done in practice. 

For instance, the ``brute force" COBE analysis of Tegmark \& Bunn (1995)
had $n=4038$ and $m=2$, and involved inverting 
$4038\times 4038$ matrices at a couple of hundred grid points.
A future high-resolution all-sky CMB map might contain 
of order $n=10^7$ pixels, and direct numerical
inversion of matrices this large is clearly unfeasible 
at the present time. Also, to numerically map out the likelihood
function in the 11-dimensional parameter space explored by 
Jungman {\etal} (1996b) with reasonable resolution
would entail evaluating it at such a large number of 
grid points, that it would be desirable to make 
the inversion at each grid point as fast as possible.

Likewise, the spherical harmonic analysis of  
the 1.2Jy redshift survey (Heavens \& Taylor, 1995, Ballinger
Heavens \& Taylor 1996), and the PSC-z survey 
(Tadros {\etal}, in preparation), 
required a likelihood analysis of
some 1208 modes to find the redshift distortion parameter and 
a nonparametric stepwise estimate of the power spectrum.

Because of this, it is common to carry out some form
of {\it data compression} whereby the $n$ data points are
reduced to some smaller set of $n'$ numbers before
the likelihood analysis is done. Since the time 
needed for a matrix inversion scales as $n^3$, even a 
modest compression factor such as $n/n'=5$ can accelerate
the parameter estimation by more than a factor of 100.

\subsection{The optimization problem}

In this section, we will discuss the special case of 
{\it linear} data compression, which has as an important
special case the so-called signal-to-noise
eigenmode method. We will return to non-linear data compression
methods below, in Section~\ref{PreCompressionSec}.

The most general linear data compression can clearly be written as
\beq{BdefEq}
\y = \B\x,
\eeq
where the $n'$-dimensional vector $\y$ is the new
compressed data set and $\B$ is some arbitrary 
$n'\times n$ matrix. 
Thus the new data set has
\beq{NewExpecEq}
\cases{
\expec{\y}				&$=\B\vmu$,\crr
\expec{\y\y^t}-\expec{\y}\expec{\y}^t	&$=\B \C \B^t$,\crr
}
\eeq
and substituting this into \eq{GaussFisherEq} we find that
the new Fisher information matrix $\Fnew$ is given by
\beqa{GaussFisherEq2}
\nonumber
\Fnew_{ij} 
&=& {1\over 2}\Tr[(\B\C\B^t)^{-1}(\B\C,_i\B^t)(\B\C\B^t)^{-1}(\B\C,_j\B^t)\\
&+&(\B\C\B^t)^{-1}(\B\M_{ij}\B^t)].
\eeqa
If $n=n'$ and $\B$ is invertible, then the $\B$-matrices cancel in
this expression, leaving us with simply 
\beq{simpleq}
\Fnew_{ij} = {1\over 2}\Tr[\B^{-t}(\A_i \A_j + \C^{-1}\M_{ij})\B^t] = \F_{ij}.
\eeq
In other words, $\B$ just makes a similarity transform of the matrix
within the trace in \eq{GaussFisherEq}, leaving the 
Fisher information matrix $\F$ unchanged.
This reflects the fact that
when $\B$ is invertible, no information is lost or
gained, so that the error bars on a measured parameter are unchanged.
For instance, replacing a galaxy-cut COBE map consisting of 
say 3600 pixels by 
its spherical harmonic coefficients up to $\l=59$ 
(there are 3600 such coefficients)
would be an invertible transformation, 
thus making no difference whatsoever as to the attainable error bars.
Likewise, expansion in galaxy-cut spherical harmonics
gives exactly the same result as expansion in an orthonormal
basis for the galaxy-cut spherical harmonics (as in G\'orski 1994), 
since the mapping from the non-orthonormal basis to the orthonormal basis 
is invertible.
This result is obvious if we think in terms of information:
if the mapping from $\x$ to $\y$ is invertible, then $\y$ must clearly contain
exactly the same information that $\x$ does, since we can reconstruct 
$\x$ by knowing $\y$. 

Rather, the interesting question is what happens when 
$n'<n$ and $\B$ is not invertible. Each row vector of $\B$ specifies 
one of the numbers in the new data set (as some linear combination 
the original data $\x$), and we wish to know which such row vectors 
capture the most information about the parameters $\vth$.
If $\B$ has merely a single row, then we can write $\B=\vb^t$ for some 
vector $\vb$, and the diagonal ($i=j$) part of \eq{GaussFisherEq2}
reduces to 
\beq{GaussFisherEq3}
\Fnew_{ii} = {1\over 2}\left({\vb^t\C,_i\vb\over\vb^t\C\vb}\right)^2 
+ {(\vb^t\vmu,_i)^2\over (\vb^t\C\vb)}.
\eeq
Let us now focus on the problem of estimating a single parameter 
$\theta_i$ when all other parameters are known --- we will return 
to the multi-parameter case below, in Section~\ref{WorldDominationSec}.
Since the error bar $\Delta\theta_i\approx \Fnew_{ii}^{-1/2}$ in 
this case, we want to find the $\vb$ that maximizes the right-hand 
side of \eq{GaussFisherEq3}.
Although this is a non-linear problem 
in general, there are two simple special cases which between them
cover many of the situations that appear in cosmology applications:
\begin{itemize}
\item The case where the mean is known: $\vmu,_i=0$
\item The case where the covariance matrix is known: $\C,_i=0$
\end{itemize}
Below we show first how these two cases can be solved separately,
then how the most general case can be for all practical
purposes solved by combining these two solutions. 

\subsection{When the mean is known}
\label{MeanIsKnownSec}

When $\vmu$ is independent of $\theta_i$, the second term in 
\eq{GaussFisherEq3} vanishes, and we simply wish to maximize the 
quantity
\beq{GaussFisherEq4}
(2\Fnew_{ii})^{1/2} = {|\vb^t\C,_i\vb|\over \vb^t\C\vb}.
\eeq
Since the mean $\vmu$ does not depend on the parameters (\ie, it is known), 
let us for simplicity redefine our data by $\x\mapsto\x-\vmu$, so that
$\expec{\x}=0$.
Note that whereas the denominator in \eq{GaussFisherEq4} is 
positive (since $\C$ is a covariance matrix and thus positive definite),
the expression $\vb^t\C,_i\vb$ in the numerator might be negative, since 
$\C,_i$ is not necessarily positive definite.
We therefore want to make $(\vb^t\C,_i\vb)/(\vb^t\C\vb)$ 
either as large as possible or as small as
possible, depending on its sign. Regardless of the sign,
we thus seek the $\vb$ for which this ratio takes an extremum, {\ie}, 
we want the derivatives with respect to the components of $\vb$ to vanish.
Since this ratio is clearly unchanged if we multiply $\vb$ by a constant, 
we can without loss of generality choose to normalize $\vb$ so that the 
denominator equals unity. 
We thus seek an extremum of $\vb^t \C,_i\vb$ subject to the constraint that
\beq{NormalizationEq}
\vb^t\C\vb = 1.
\eeq
This optimization problem is readily solved by introducing a Lagrange 
multiplier $\lambda$ and extremizing the Lagrangean
\beq{LagrangeEq}
\vb^t\C,_i\vb - \lambda\vb^t\C\vb.
\eeq
Varying $\vb$ and setting the result equal to zero, we obtain
the generalized eigenvalue problem
\beq{EigenEq}
\C,_i\vb = \lambda \C\vb.
\eeq
Since $\C$ is symmetric and positive definite, 
we can Cholesky decompose it (\eg, Press {\etal} 1992) as 
$\C=\bL\bL^t$ for some invertible matrix $\bL$.
Multiplying \eq{EigenEq} by $\bL^{-1}$ from the left, we 
can rewrite it as 
\beq{EigenEq2}
(\bL^{-1}\C,_i\bL^{-t})(\bL^t\vb) = \lambda (\bL^t\vb),
\eeq
which we recognize as an ordinary eigenvalue problem for the 
symmetric matrix $(\bL^{-1}\C,_i\bL^{-t})$,
whose solution will be $n$ orthogonal eigenvectors 
$(\bL^t\vb_k)$ and corresponding eigenvalues $\lambda_k$,
$k=1,2,...,n$.
Let us sort them so that 
\beq{SortEq}
|\lambda_1|\geq|\lambda_2|\geq...\geq|\lambda_{n}|.
\eeq
Let us choose the $\kth$ row of $\B$ to be the row vector 
$\vb_k^t$, so that the compressed data set is given by
$y_k = \vb_k^t\x$.
The orthogonality property 
$(\bL^t\vb_k)\cdot(\bL^t\vb_{k'})\propto\delta_{kk'}$
in combination with our chosen normalization of \eq{NormalizationEq} 
then tells us that our compressed data satisfies
\beq{UncorrEq}
\expec{y_k y_{k'}} = \expec{(\vb_k^t\x)(\x^t\vb_{k'})} = 
\vb_k^t \C \vb_{k'} = \vb_k^t \bL\bL^t \vb_{k'} = \delta_{kk'},
\eeq 
{\ie}, $\expec{\y\y^t}=\I$.
The compressed data values $\y$ thus have the nice
property that they are what is known as 
{\it statistically orthonormal}, {\ie}, 
they are all uncorrelated and all have unit variance.
Since we are assuming that everything is Gaussian, this 
also implies that they are statistically independent ---
in fact, $\y$ is merely a vector of independent unit
Gaussian random variables. 
In other words, knowledge of $y_k$ gives us no information
at all about the other $y$-values. 
This means that the entire information content of the initial 
data set $\x$ has been portioned out in disjoint pieces in the
new compressed data set, where $y_1$ contains the most information
about $\theta_i$, $y_2$ is the runner up
(once the information content of $y_1$ has been removed, 
$y_2$ contains the most information
about $\theta_i$), and so on.

If we use all the $n$ vectors $\vb_k$ as rows in $\B$, 
then $\B$ will be invertible, and we clearly have not achieved 
any compression at all. However, once this matrix $\B$ has been found,
the compression prescription is obvious: if we want a compressed data
set of size $n'<n$, then we simply throw away all but the
first $n'$ rows of $\B$. 
It is straightforward to show 
(see {\eg} Therrien 1992; Vogeley \& Szalay 1996)
that if we fix an integer $n'$
and then attempt to minimize $\Delta\theta_i$, we will
arrive at exactly the same $\B$ as with our prescription above
(or this $\B$ multiplied by some 
invertible matrix, which as we saw will leave the
information content unchanged). 
   
How does the error bar $\Delta\theta_i$ depend on the choice 
of $n'$? \Eq{EigenEq} implies that 
\beq{MatrixEigenEq2}
\C,_i\B^t = \C\B^t\bLambda,
\eeq
where $\Lambda_{ij}\equiv\delta_{ij}\lambda_i$, {\ie}, a
diagonal matrix containing all the eigenvalues.
Since \eq{UncorrEq} implies that
\beq{MatrixEigenEq2a}
\B\C\B^t = \I,
\eeq
it follows from \eq{MatrixEigenEq2} that
\beq{MatrixEigenEq2b}
\B\C,_i\B^t = \bLambda.
\eeq
Therefore the 
diagonal part of \eq{GaussFisherEq2} reduces to simply
\beq{DiagFisherEq}
2\Fnew_{ii} = \Tr[\{(\B\C\B^t)^{-1}(\B\C,_i\B^t)\}^2] 
=\Tr\bLambda^2 = \sum_{k=1}^{n'} \lambda_k^2.
\eeq
It is thus convenient 
to plot $\lambda_k^2$ as a function of the mode number $k$,
as we have done in \fig{BetaLambdaFig} for the specific case
of measuring the redshift distortion parameter from the
 1.2Jy redshift survey, to see at what $k$ the information 
content starts petering out. Alternatively,
one can plot $\Delta\theta_i$ as a function of $n'$
as in \fig{DeltaBetaFig} and see when the curve
flattens out, {\ie}, when the computational
cost of including more modes begins to yield diminishing 
returns.
If one feels uncomfortable about choosing 
$n'$ by ``eyeballing", one can obtain a more quantitative
criterion by noting that using only a single
mode $\vb_k$ would give $\Delta\theta_i = 1/|\lambda_k|$. 
Thus the quantity $\theta_i|\lambda_k|$ can be thought of as 
the signal-to-noise ratio for the $\kth$ mode, and one 
can choose to keep only those modes where this ratio exceed
unity (or some other constant such as $0.2$).

\subsection{When the covariance matrix is known}
\label{CovIsKnownSec}

When $C$ is independent of $\theta_i$, the first term in 
\eq{GaussFisherEq3} vanishes, and we simply wish to maximize the 
quantity
\beq{GaussFisherEq5}
\Fnew_{ii} = {\vb^t\M_{ii}\vb\over \vb^t\C\vb}.
\eeq
Introducing a Lagrange multiplier $\lambda$ and proceeding 
exactly as in the previous section, 
this is equivalent to the eigenvalue problem
\beq{EigenEq3}
(\bL^{-1}\M_{ii}\bL^{-t})(\bL^t\vb) = \lambda (\bL^t\vb).
\eeq
However, since the matrix $\M_{ii}=2\vmu,_i\vmu,_i^t$ is
merely of rank one, this problem is much simpler than that 
in the previous section. Since the left hand side is
\beq{EigenEq4}
2(\bL^{-1}\vmu,_i)(\vmu,_i^t\vb) \propto (\bL^{-1}\vmu,_i),
\eeq
it points in the direction $(\bL^{-1}\vmu,_i)$ regardless what the
vector $\vb$ is, so the only non-trivial 
eigenvector of $(\bL^{-1}\M_{ii}\bL^{-t})$
is 
\beq{EigenEq5}
(\bL^t\vb_1) = (\bL^{-1}\vmu,_i),
\eeq
with a corresponding eigenvalue 
\beq{EigenEq6}
\lambda_1 = 2|\bL^{-1}\vmu,_i|^2 = \Tr[\C\M_{ii}].
\eeq
All other eigenvalues vanish.
In other words, $\vb_1 = \C^{-1}\vmu,_i$,
and the new compressed data set consists of just one single 
number, 
\beq{b1eq}
y_1 = \vmu,_i^t \C^{-1}\x
\eeq
which contains just as much information about 
the parameter $\theta_i$ as the entire 
data set $\x$ did.
Another way to see this is to compute the 
Fisher information before and after the compression, 
and note that
\beq{TrivialFisherEq}
\Fnew_{ii} = \F_{ii} = \vmu,_i^t \C^{-1}\vmu,_i.
\eeq
In other words, all the information about 
$\theta_i$ in $\x$ is
has been distilled into the number $y_1$. 
Thus the ML-estimate of $\theta_i$
is just some function of $y_1$. 
For the special case where $\vmu$ depends linearly on
the parameter, {\ie}, $\vmu = \vmu,_i\theta_i$ where 
$\vmu,_i$ is a known constant vector,
this function becomes trivial:
the expectation value of \eq{b1eq} reduces to
\beq{TerminalCompressionEq}
\expec{y_1} = (\vmu,_i^t \C^{-1}\vmu,_i)\,\theta_i,
\eeq
which means that on average, up to a 
known constant $(\vmu,_i^t \C^{-1}\vmu,_i)$, 
the compressed data is just 
the desired parameter itself.

\subsection{When neither is known --- the general case}

In the most general case, both $\vmu$ and $\C$ depend on the
parameters $\vth$. This happens for instance in the 
CMB case when the components of $\x$ are chosen to be estimates
of the power spectrum $C_\l$, as in   
Section~\ref{PreCompressionSec} below.

Clearly, the more ways in which the probability distribution 
for $\x$ depends on the parameters, the more accurately 
we can estimate them. In other words, when both $\vmu$ and $\C$
depend on $\vth$, we expect to do even better than in the two cases
discussed above.
Finding the vectors $\vb$ that extremize  
\eq{GaussFisherEq3} is quite a difficult numerical problem.
Fortunately, this is completely unnecessary.
Solving the eigenvalue problem in 
Section~\ref{MeanIsKnownSec} gives us $n''$ 
compression vectors capturing the bulk of the cosmological 
information coming from the first term in \eq{GaussFisherEq3}.
Section~\ref{CovIsKnownSec} gives us one additional
vector $\vmu,_i^t\C^{-1}$ that captures all the cosmological 
information coming from the second term.
In other words, if we simply use all these $n''+1$ 
compression vectors
together, we have for all practical purposes solved our 
problem. Since a direct numerical attack on \eq{GaussFisherEq3}
could never reduce these $n''+1$ vectors to fewer than $n''$
without widening the resulting error bars, the
time savings in the ensuing likelihood analysis
would at most be a completely 
negligible factor $[(n''+1)/n'']^3 \sim 1 + 3/n''$
compared the simple prescription we have given here.

\subsection{Estimating several parameters at once}
\label{WorldDominationSec}
 
The Karhunen-Lo\`eve method described above 
was designed to condense the bulk of the 
information about a {\it single} parameter into as 
few numbers $y_k$ as possible. Although this particular 
problem does occasionally arise, most cosmology applications 
are more complicated, involving models that contain more
than one unknown parameter. The previous applications 
of Karhunen-Lo\`eve methods to CMB maps
(Bond 1994; Bunn \& Sugiyama 1995; Bunn 1995) and 
galaxy surveys (Vogeley 1995; Vogeley \& Szalay 1996), have 
involved taking a rather minimalistic approach to this issue,
by optimizing the eigenmodes for one particular
parameter (typically the overall normalization of the
fluctuations) and assuming that these eigenmodes will 
capture most of the relevant information about the 
other parameters as well.

If we want to estimate several parameters from a data set 
simultaneously, then how should we best compress the data?
As we saw in Section~\ref{FisherSec}, the 
covariance matrix $\T$ of our parameter estimates $\vth$ 
is approximately $\F^{-1}$, so for the multivariate case,
we have
\beq{MultivarSdevEq}
\Delta\theta_i = (\F^{-1})_{ii}^{1/2}
\eeq
instead of $\Delta\theta_i = \F_{ii}^{-1/2}$.
One approach to the problem would thus be trying to 
minimize the diagonal elements of $\F^{-1}$ rather than,
as above, trying to minimize the diagonal elements of $\F$.
This is, however, quite a cumbersome optimization problem.

Fortunately, there is an alternative solution to the problem
which is easy to implement in practice and 
in addition is easy to understand intuitively,
in terms of information content.
Suppose that we repeat the standard KL procedure 
described above $m$ times, once for each parameter $\theta_i$, 
and let $n'_i$ denote the number of modes that we choose to use
when estimating the $\ith$ parameter.
We then end up with $m$ different compressed data sets, 
such that the $n'_i$ numbers in the $\ith$ set 
contain essentially all the information that 
there was about $\theta_i$.
This means that the union $\y$ of all these compressed 
data sets, consisting of 
\beq{nSumEq}
n''=\sum_{i=1}^m n_i'
\eeq
numbers,
retains basically all the information from 
$\x$ that is relevant for our joint estimation
of all the parameters.
The problem is of course that there 
might be plenty of redundancy in $\y$. 
Indeed, if we typically compressed by say a 
factor $n/n''_i\sim 10$ and had  
11 parameters as in the CMB model of 
Jungman {\etal} (1996b), then we would have 
achieved a counterproductive 
``anti-compression" with $n''>n$.
 
How can we remove the unnecessary ``pork" from $\y$?
The $n'_i$ eigenvectors selected via the
KL compression method 
span a certain subspace of the $n$-dimensional 
space in which the data vector $\x$ lives, and we can think 
of the data compression step as simply projecting 
the vector $\x$ onto this subspace.
For each parameter $\theta_i$, we found such a subspace,
and the redundancy stems from the fact that many of these 
subspaces overlap with one another, or point in very
similar directions.
In order to compress efficiently and yet retain almost all the 
the information about all the parameters,
we want to find a small set of vectors that span as much as possible 
of the union of all the subspaces.
As described in detail in {\eg} Press {\etal} (1992),
this is readily accomplished by making a
{\it Singular Value Decomposition} (SVD)
and throwing away all vectors corresponding to
very small singular values. 
Let us define $\B_i=\B\bLambda$, where $\B$ is 
the KL compression matrix optimized for 
estimating the $\ith$ parameter. Here we multiply the row vectors 
by their corresponding eigenvalues so that better vectors will 
receive more weight in what follows.
Combining all the row vectors of the transposed compression
matrices $\B_i^t$ into a single matrix, 
SVD allows us to factor this matrix as 
\beq{SVDeq}
\left(\B_1^t\>\dots\>\B_m^t\right) = \U\bLambda\V^t,
\eeq
where $\U^t\U=\I$, $\V^t\V=\I$ and the matrix $\bLambda$
is diagonal. One customarily writes $\bLambda=\diag\{\lambda_i\}$,
and refers to the diagonal elements $\lambda_i$ 
as the {\it singular values}. These are basically 
a generalization of eigenvalues to non-square matrices.
We define our final compression matrix as
\beq{SVDBeq}
\B\equiv\U^t.
\eeq
The columns of $\U$
with corresponding non-zero singular values $\lambda_i$ 
form an orthonormal set of basis
vectors that span the same space as all the initial 
compression vectors together.
The column vectors
of $\U$ with vanishing singular values form an orthonormal
basis for the null-space of $\U\bLambda\V^t$,
representing redundant information.
Similarly, the vectors corresponding to singular 
values near zero capture information that 
is almost entirely redundant.
By making a plot similar to \fig{BetaLambdaFig},
showing $\lambda_k$ as a function of $k$, 
one decides on the final number $n'<n''$ of 
row vectors in $\B$ to keep.
Once $n'$ has been fixed, one can of course 
(if one prefers\footnote{
The only merit of the statistical orthogonality
is that it will make $\expec{\y\y^t}$ more close to diagonal  
near the fiducial parameter values $\vth$, which might
conceivably speed up the 
matrix inversion if an iterative method is used.
}) make the $n'$ numbers
in the new compressed data set statistically orthogonal
as before by Cholesky decomposing their covariance matrix 
$\B\C\B^t=\bL\bL^t$ and replacing $\z=\B\x$ by $\bL^{-t}\z$.
In summary, we have found that when we wish to estimate more
than one parameter from the data, 
we can obtain a close to optimal data compression  
with the following steps:
\begin{enumerate}
\item Compute the KL eigenmodes that contain the bulk of the information
about the first parameter.
\item Repeat this procedure separately for all other parameters.
\item Arrange all the resulting vectors, multiplied by their eigenvalues, 
as the rows of $\B$.
\item Make an SVD of $\B$ and throw away all rows corresponding to
very small singular values.
\end{enumerate}

\subsection{The problem of not knowing the parameters a priori}

The KL-approach has sometimes been criticized for not being
model-independent, in the sense that one needs to make an a priori
guess as to what the true parameter values $\vth_0$ are
in order to be able to compute the compression matrix $\B$.
Although there is no evidence that a bad guess as to $\vth_0$ 
biases the final results (see {\eg} Bunn 1995 for a detailed discussion
of these issues, including numerical examples), 
a poor guess of $\vth_0$ will of course lead to a data compression
that is no longer optimal.
In other words, one would expect to still get an unbiased 
answer, but perhaps with larger error bars than necessary.

In practice, this loss of efficiency is likely to 
be marginal. For instance, as a test, Bunn (1995) performed a KL-analysis of
a COBE CMB map with $n\sim 1$ assuming a blatantly erroneous
spectral index $n=2$ to compute $B$
($n=2$ is ruled out at about $3\sigma$ a posteriori),
and compared this with the results obtained when assuming $n=1$.
The error bars where found to change only marginally.

If one nonetheless wishes to do something about this efficiency problem,
it is fortunately quite easy in practice.
The simplest approach is an iterative scheme, whereby the KL-analysis 
is carried out twice, using the ML-estimate of $\vth$ from the first 
run as the fiducial ``true" value in the second run. 
A more rigorous approach is to compute compression vectors $\vb$ 
for a number of different assumptions about $\vth_0$ that 
include the extreme corners of parameter space, and then combine
all these vectors into a single compression matrix $\B$ by singular-value 
decomposition exactly as described in the previous section.

\section{COSMOLOGY APPLICATIONS}
\label{ApplicationsSec}

In this section, we illustrate the theoretical discussion above with 
a number of cosmology examples, and show how the recently published 
work on CMB maps and galaxy surveys fits into the general KL-scheme 
as special cases. We will see that the typical compression factor 
is about 10 in both an IRAS example and a COBE example.

\subsection{A large-scale-structure example: redshift-space distortions}
\label{BetaSec}

Here we apply a KL-analysis to the problem
of redshift-space distortions in the 1.2Jy IRAS galaxy 
survey (Fisher {\etal} 1995). 
This problem has previously been analyzed in 
detail by Heavens \& Taylor (1995, hereafter HT95) who used a spherical
harmonic analysis. 
Here we repeat their analysis, but
include a KL data-compression step to investigate how many 
modes are needed for an accurate determination of the 
redshift distortion parameter
\beq{BetaDefEq}
\beta\equiv {\Omega^{0.6}\over b},
\eeq
where $b$ denotes the conventional linear bias factor, 
the ratio between the fluctuations in luminous matter 
and the total matter fluctuations.
As was first shown by Kaiser (1987),
the peculiar motions of galaxies induces a 
characteristic radial distortion in the apparent 
clustering pattern that depends only on this 
parameter $\beta$.

\begin{figure}
\epsfxsize=9.cm
\epsfysize=13.cm
\vspace{-3.cm}
\hfill
\epsfbox{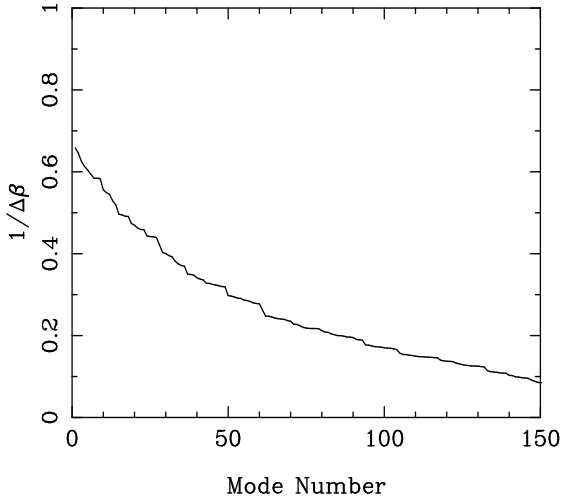 }
\hfill
\epsfverbosetrue
\vspace{-3.6cm}
\caption{KL-eigenvalues $\lambda=1/\Delta\beta$. }
\label{BetaLambdaFig}
\end{figure}

\begin{figure}
\epsfxsize=9.cm
\epsfysize=13.cm
\vspace{-3.cm}
\hfill
\epsfbox{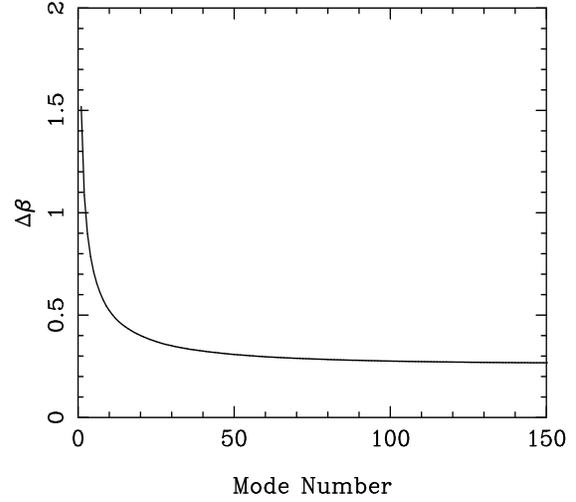 }
\hfill
\epsfverbosetrue
\vspace{-3.6cm}
  \caption{Error bar on beta $\beta$ as a function of $n'$, the
  number of eigenmodes used.}
\label{DeltaBetaFig}
\end{figure}

As the initial data vector $\x$, we 
use the coefficients obtained by expanding
the observed galaxy distribution in 
combinations of spherical harmonics and spherical Bessel functions
as described in HT95, with the known mean values subtracted off
so that $\vmu=\expec{\x}=0$.
HT95 show that the covariance matrix is given by 
\beq{BetaCovEq}
\C_{\mu\nu} = \Lambda^{\rm sn}_{\mu\nu}  + 
{1\over 2}\sum_\alpha(\bPhi_{\alpha\mu}
+\beta\V_{\alpha\mu})(\bPhi_{\alpha\nu}+\beta\V_{\alpha\nu})P(k_\alpha), 
\eeq
where the indices $\mu$, $\nu$ and $\alpha$ 
run over the above-mentioned modes, 
and $P(k)$ is the power spectrum.
The matrices $\bPhi$ and $\V$ encapsulate the effects of finite 
survey volume convolution and 
redshift distortions, respectively.
The matrix $\bLambda^{\rm sn}$ contains the shot-noise contribution.  
All three matrices are independent of $\beta$.
For our illustrative example, we assume a parametrised form of the
power spectrum suggested by Peacock \& Dodds (1994), 
which means that $\beta$ is the only unknown parameter.
Hence $m=1$ and $\vth=\theta_1=\beta$.
Differentiating \eq{BetaCovEq}, we obtain 
\beq{BetaCprimeEq}
\C,_1 = {\partial \C\over\partial\beta} = 
{1\over 2}(\V \bP \bPhi^t+\bPhi \bP \V^t) + (\V \bP \V^t)\beta,
\eeq
where $\bP_{\alpha \beta}\equiv P(k_{\alpha}) \delta_{\alpha \beta}$
(\ie, $\bP=\diag\{P(k_{\alpha}\}$).
We assume an a priori value $\beta=1$ to evaluate this matrix.
Using $n=1208$ modes in the initial data set $\x$ as in HT95,
we obtain the eigenvalues $\lambda_k$ shown in \fig{BetaLambdaFig}.
The resulting error bar 
\beq{DeltaBetaEq}
\Delta\beta = \left[\sum_{k=1}^{n'} \lambda_k^2\right]^{-1/2}
\eeq
is plotted as a function of $n'$ (the number of modes 
used) in \fig{DeltaBetaFig}, and is seen to level out at around
$n'=100$. 
This means that although HT95 
used $\sim 10^3$ modes to estimate $\beta$, almost all the information 
is actually contained in the best $\sim 10^2$ linear combinations 
of these modes.
In other words, we can obtain basically identical results to those
found in HT95 by using a compressed data set only $10\%$ of the original size.
Since the matrix inversion time scales as $n'^3$, this compression
factor of 10 thus allows the inversion to be
carried out 1000 times faster.

\subsection{A special case: the signal-to-noise eigenmode method}

The above-mentioned example was rather generic in the sense that 
$\C$ depended on $\vth$ in a nonlinear way 
(in this case, $\C$ depended quadratically on $\beta$).
However, the special case of the KL-method where the parameter 
dependence is linear (or rather affine)
is so common and important that it has acquired a special name: 
the {\it signal-to-noise eigenmode method}.
This is the special case where $\mu=0$, we have only one
unknown parameter $\theta$, and the covariance matrix can be written in
the form 
\beq{SNcovEq}
\C = \Sig \theta + \N,
\eeq
where the known matrices $\Sig$ and $\N$ are independent of $\theta$.
For reasons that will become clear from the examples below, 
$\Sig$ and $\N$ are normally referred to as the {\it signal} and 
{\it noise} matrices, respectively, and they are normally both positive
definite. 
Since $d\C/d\theta = \Sig$,  
\eq{EigenEq} gives the generalized eigenvalue problem
\beq{SNeigenEq}
\Sig\vb = \lambda (\Sig+\N)\vb.
\eeq
By Cholesky decomposing the noise matrix as $\N=\bL\bL^t$, 
this is readily rearranged into the ordinary 
eigenvalue problem 
\beq{SNeigenEq2}
(\bL^{-1}\Sig\bL^{-t})(\bL^t\vb) = \lambda'(\bL^t\vb),
\eeq
where $\lambda'\equiv\lambda/(1-\lambda)$.
Since the matrix to be diagonalized, $(\bL^{-1}\Sig\bL^{-t})$, 
is loosely speaking $(\N^{-1/2}\Sig\N^{-1/2})$, a type of signal-to-noise
ratio, its eigenvectors $\vb$ are usually referred to as the
{\it signal-to-noise eigenmodes}. Historically, this terminology 
probably arose because one was analyzing one-dimensional time series 
with white noise, where $\N\propto \I$, 
so that $(\bL^{-1}\Sig\bL^{-t}) = \Sig\N^{-1}$ was the signal-to-noise ratio
even in the strict sense.
Vogeley \& Szalay (1996) refer to the change of variables 
$\x\mapsto \bL^{-1}\x$ as {\it prewhitening}, since it transforms 
$\Sig+\N$ into $(\bL^{-1}\Sig\bL^{-t}) + \I$, {\ie}, 
transforms the noise matrix 
into the white noise matrix, the identity matrix.

\Eq{UncorrEq} showed that in the general KL-case, 
the compressed data set was statistically orthonormal, 
$\expec{\y\y^t}=\I$.
In the S/N-case, the compressed data has an additional 
useful property: both the signal and noise 
contributions to $\y$ are uncorrelated {\it separately}.
It is easy to show that 
\beq{SNorthoEq}
\vb_k^t\Sig\vb_{k'}  = \lambda' \vb_k^t\N\vb_{k'} \propto \delta_{kk'},
\eeq
which means that the covariance matrix $\expec{\y\y^t}$
will remain diagonal even if we have misestimated
the true signal-to-noise ratio. In other words, 
if the sole purpose of our likelihood analysis
is to estimate the overall normalization $\theta$
from \eq{SNcovEq}, we only need to invert 
diagonal matrices.

The signal-to-noise method arises in a wide variety of contexts.
For instance, it is a special case not only of the 
KL-method, but also of the power spectrum estimation method
of Tegmark (1996), corresponding to the case were the width of the
window functions is ignored.

\subsubsection{Signal-to-noise analysis of CMB maps}
\label{cmbSNsec}

The signal-to-noise eigenmode method was introduced into CMB 
analysis by Bond (1994), who applied it to sky maps from 
both the COBE and FIRS experiments. It has also been applied
to the COBE data by 
Bunn\footnote{
In fact, both Bond and Bunn independently reinvented the entire KL-method.
}, and used it to constrain a wide variety
of cosmological models (Bunn 1995; Bunn \& Sugiyama 1995;
Bunn, Scott \& White 1995; Hu, Bunn \& Sugiyama 1995; White \& Bunn 1995;
Yamamoto \& Bunn 1996). 
Here the uncompressed data set consists of the CMB 
temperatures in $n$ pixels from a sky map (for instance, 
$n=4038$ or $4016$ for a COBE map with a $20^\circ$ galactic cut, depending on the
pixelization scheme used). If foreground contamination is negligible
(see {\eg} Tegmark \& Efstathiou 1996 for a recent review) 
and the pixel noise 
is uncorrelated (as it is for COBE to an excellent 
approximation --- see Lineweaver {\etal} 1994) one has 
\beq{CMBcorrEq}
\cases{
\mu	&$=0$,\crr
\Sig_{ij}	&$=\sum_{\l=2}^\infty 
	\left({2\l+1\over 4\pi}\right) P_\l(\nh_i\cdot\nh_j) 
 	W_\l^2 C_\l,$\crr
\N_{ij}	&$=\delta_{ij}\sigma_i^2,$
}
\eeq 
where $\nh_i$ is a unit vector pointing the direction of the 
$\ith$ pixel, $\sigma_i$ is the {\rms} noise in this pixel, 
$P_\l$ are the Legendre polynomials and $W_\l$ is the 
experimental beam function.  
This expression for $\Sig$ should only be used if one 
marginalizes over the monopole and dipole in
the ensuing likelihood analysis --- otherwise $\Sig$ should be 
corrected as described 
in Tegmark \& Bunn (1995). 
In all the above-mentioned applications, the compression was optimized 
with respect to the overall normalization of the power spectrum 
(say $\theta=C_2$), so the matrix $\Sig$ was independent of $\theta$. 
It has been found (Bunn \& Sugiyama 1995) that a compression 
factor of about 10, down to $n'\sim 400$ modes, can be attained without significant 
loss of information. This is about the same compression factor that we
obtained in our redshift distortion example above.

Although these above-mentioned papers constrained a wide variety
of cosmological parameters, 
the compression was in all cases optimized only for one parameter, the
overall power spectrum normalization.
Although this procedure can be improved as was described
in Section~\ref{WorldDominationSec}, 
this does not appear to be causing substantial loss of efficiency
in the COBE case.
Support for this conclusion
is provided by Tegmark \& Bunn (1995), where it is found that 
KL-compression optimized for measuring the normalization 
gives error bars on the spectral index $n$ that are less than 
$10\%$ away from the minimal value that is obtained without 
any compression at all.
It should also be noted that if one optimized the KL-compression for 
a different parameter $\theta$, say $\theta=n$, then 
$\C$ would no longer be of the simple form of \eq{SNcovEq}.
Thus the signal-to-noise eigenmode treatment no longer applies, and the
more general \eq{EigenEq} must be used instead.

As has been pointed out by G\'orski (1994), the rapid fall-off of the 
COBE window function $W_\l$ implies that virtually all 
the cosmological information
in the COBE maps is contained in the multipoles $\l\leq 30$. 
When implementing 
the KL-compression in practice, it is useful to take advantage of this 
fact by replacing the data set by the $957$ spherical harmonic coefficients
$a_{\l m}$ for $2\leq\l\leq 30$, since this reduces the size of the matrix
to be diagonalized by about a factor $4$.
This is an example of what we will call {\it pre-compression}, 
and we will return
to this topic below, in Section~\ref{PreCompressionSec}.

\subsubsection{Signal-to-noise analysis of galaxy surveys}

The application of the signal-to-noise eigenmode method to galaxy surveys 
was pioneered by Vogeley (1995) and has since been further elaborated
by Vogeley \& Szalay (1996). 
Although these are both method papers, deferring actual parameter
estimation to future work, the former explicitly evaluates the  
eigenmodes for the CfA survey and shows that there are about
$10^3$ modes with a signal-to-noise ratio exceeding unity.

In galaxy survey applications, the noise matrix $\N$ contains 
the contribution from Poisson shot 
noise due to the finite number of galaxies per unit volume, 
rather than on detector noise as in the CMB case.
With galaxies, the question of what to use as the initial data vector 
$\x$ is not as simple as in the CMB case, since there is 
no obviously superior way to ``pixelize" the observed density field.
Vogeley \& Szalay (1996) divide space into a large number of 
disjoint volumes (``cells"),
and derive the resulting signal matrix $\Sig$ 
in the approximation of ignoring redshift distortions.
One alternative is using ``fuzzy" cells as discussed 
by Tegmark (1995), 
for instance averages of the galaxy distribution where the 
weight functions are Gaussians centered at some grid of points.

Another alternative, which has the advantage of making it easier
to include the effect of redshift distortions, 
is to use the coefficients from an expansion in 
spherical harmonics and spherical Bessel functions as 
was done by HT95, Ballinger {\etal} (1995; hereafter BHT95)
and in Section~\ref{BetaSec} above. In BHT95 the real space power
spectrum of density perturbations was left as a free
function, to be evaluated in a stepwise fashion, along with 
the distortion parameter. In the case of only estimating
the power, by fixing $\beta$ to some constant value,
the problem reduces to a signal-to-noise eigenmode problem,
but with $m$ free parameters --- the power at 
$m$ specified wavenumbers. Due to the mixing of wavenumbers
by the $\bPhi$ and $\V$ matrices, these are generally not
independent and so we must apply the SVD method to construct
orthogonal eigenmodes. This application will be discussed elsewhere.

We conclude this section with a few additional comments on the 
above-mentioned ``pixelization" issue.
Compared to a CMB analysis, a galaxy survey analysis 
involves one more step,
which is reducing the point data to the ``initial" data 
vector $\x$.
What is the relation between the first step (weighting galaxies)
and the second step (weighting modes)?
In Feldman, Kaiser \& Peacock (1994) and Heavens \& Taylor (1995),
schemes for optimal weighting of galaxy data were presented. This 
weighting of the data (step 1) is prior to and 
complementary to 
the mode-weighting discussed in this paper (step 2). 
The data-weighting is designed
to maximize the signal-to-noise of individual modes, and one might
intuitively expect thus this to
be a good mechanism for obtaining large eigenvalues $\lambda_k$.   
The techniques of the present paper can then subsequently be used to 
trim the data in an optimal way. The data-weighting scheme
alone does not guarantee that we get the best
answer possible. This is because it maximizes the diagonal 
elements in the covariance matrix, ignoring correlations between
modes. This does therefore not guarantee 
that the complete Fisher information 
matrix will be optimal. In general, it does not 
seem tractable to devise a 
data-weighting scheme which optimizes the Fisher matrix 
directly.

\subsection{A multi-parameter CMB example}
\label{SVDsec}

Here we apply a KL-analysis to the problem of estimating 
the quadrupole normalization $Q$, the spectral index $n$ and
the reionization optical depth $\tau$ from the 4-year COBE data
(Bennett {\etal} 1996). As the initial data vector $\x$, we use the 
temperatures in the $N=4016$ pixels that lie more than 
$20^\circ$ from the galactic plane. 
Just as in Section~\ref{cmbSNsec}, $\vmu=0$ and 
$\C=\Sig+\N$ is given by \eq{CMBcorrEq}. Computation of 
$\C,_i$ thus reduces to computing the derivatives of the
angular power spectrum $\Cl$ that occur in the sum, and 
we do this in practice using the software described in 
Seljak \& Zaldarriaga (1996). The fiducial model has the 
standard CDM power spectrum shown in 
\fig{DerivativesFig}.
\begin{figure}
\centerline{\rotate[r]{\vbox{\epsfysize=9cm\epsfbox{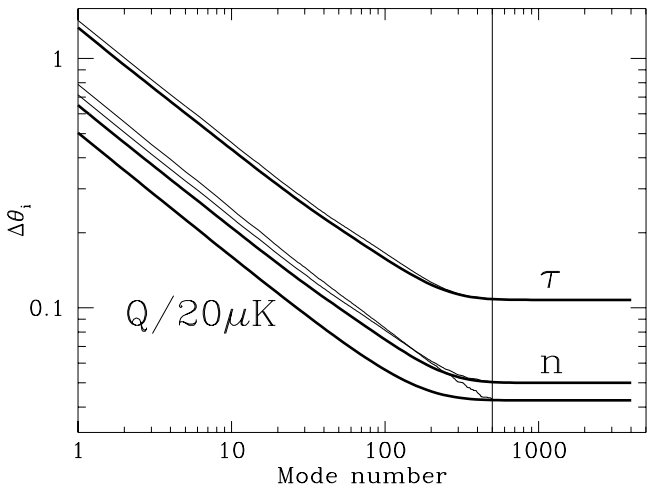}}}}
\caption{
\label{SVDfig}
The three heavy lines show the smallest error bars attainable for
the three parameters
$Q$, $n$ and $\tau$ as a function of the number of modes used, 
i.e., the error bars obtained when using the 
separately optimized KL-modes for each parameter.
The three thin lines show the error bars obtained when using
the 500 SVD modes, illustrating
that these contain essentially all the relevant information about
all three parameters.
}
\end{figure}
The three heavy lines in \fig{SVDfig} show the results of 
performing a separate KL-analysis for each parameter.
In good agreement with the findings of 
Bunn \& Sugiyama (1995), we see that virtually all information
about $Q$ is contained in the first 400 modes.
Similarly, we see that all essentially all the
information about $n$ and $\tau$ is contained in the first
400 modes optimized for these parameters.
Since the KL-modes for a parameter by construction give
the smallest error bars, the curves corresponding to
any other choice of modes would lie above these solid curves.
For instance, if we used the KL-modes optimized for 
$\tau$ to measure $Q$, the resulting curve would lie above the 
bottom one if we were to plot it in \fig{SVDfig}.
\Fig{DerivativesFig} provides a simple interpretation of this:
since $dC_\l/d\tau\approx 0$ for $\l\ll 10$, 
the $\tau$-modes do not contain information about 
the lowest multipoles, since this information is useless 
for measuring $\tau$
(even though it is important for measuring $Q$). 

We implement the SVD technique described in 
Section~\ref{WorldDominationSec} by taking the 500 best modes
for each parameter and SVD-compressing these down 
to 500 modes.
The thin lines in \fig{SVDfig} show the resulting error bars.
We see that although they lie above the minimal curves
for $k\ll 500$, they all ``catch up" at $k=500$. 
In other words, these 500 modes retain virtually
all the relevant information about $Q$, $n$ and $\tau$, since
using all of them gives virtually identical error bars 
to those obtained when using the full $N=4016$ uncompressed
data set.

\subsection{A low noise CMB example}
\label{LowNoiseSec}

\begin{figure}
\centerline{\rotate[r]{\vbox{\epsfysize=9cm\epsfbox{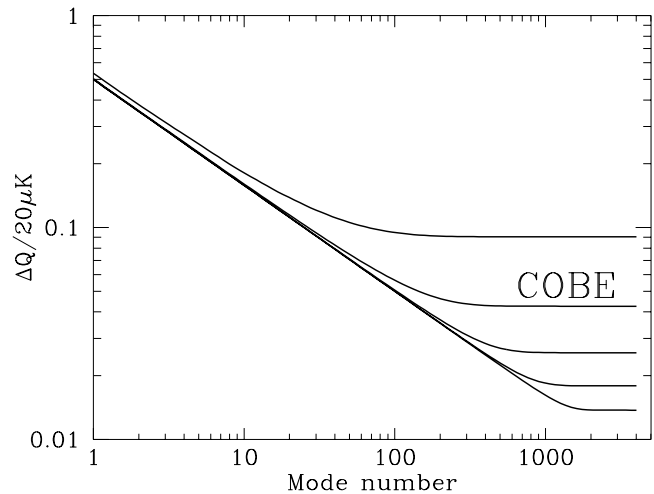}}}}
\caption{
\label{SNfig}
The error bars on the power spectrum normalization are shown for 
hypothetical COBE experiments with different noise levels.
From top to bottom, they correspond to a noise enhancement
factor 10, the real 4 year data, and noise reduction factors of
10, 100 and 1000.
}
\end{figure}

How effective will KL-compression be for future CMB data sets?
Might it be the case that when noise levels are much lower, then 
all $N$ KL-modes will have S/N$\gg 1$, so that no compression
is possible?
\Fig{SNfig} shows that the answer to the second question is no.
For this example, we have repeated the above-mentioned COBE analysis
for a range of different noise levels. 
With ten times less noise, the compression factor is seen to drop
to about $4016/700\sim 7$ and another order of magnitude of noise 
reduction lowers the compression factor to about 4.
The upcoming MAP experiment forecasts a noise 
level $w^{-1}\equiv 4\pi \sigma^2/N\approx 2.5\times 10^{-15}$, 
as compared to $w^{-1}\approx 10^{-12}$ for COBE and
$w^{-1}\approx 5\times 10^{-18}$ for the planned COBRAS/SAMBA experiment.
These are quadratic quantities, so taking square roots, we see
that MAP and COBRAS/SAMBA reduce the COBE noise by factors around
20 and 450, respectively.
\Fig{SNfig} shows that even with 1000 times less noise, 
a compression factor of 2 is readily attainable.
The explanation of this is oversampling. To avoid aliasing problems, the
mean pixel separation must be smaller than the beam width by a substantial
factor (the Shannon oversampling factor $\sim$ 2.5). 
This redundancy remains even when the data set itself has excellent 
signal-to-noise, so the KL-compression can take advantage 
of the fact that
the pixel basis of the map is a sense overcomplete. 
Although future CMB experiments will of course have a much higher angular
resolution than in our example, the oversampling factor
will remain similar, so our conclusion about the usefulness of 
KL-compression will remain the same.

\section{ON GIANT DATA SETS: THE NEED FOR PRE-COMPRESSION}
\label{PreCompressionSec}

Above we have shown that the KL-compression technique is in many cases
very useful. In this section, we will discuss some of its limitations,
as well as outline nonlinear extensions that may make the 
compression technique feasible even for the next generation of 
CMB missions and galaxy surveys.

The number of pixels in a sky map from a next-generation CMB mission
is likely to be around $10^7$. The Sloan Digital Sky Survey (SDSS) will 
measure the redshifts of $10^6$ galaxies.
Is it really feasible to apply KL-methods and likelihood analysis to
data sets of this gargantuan size?
We will argue that although an orthodox KL-analysis is {\it not}
feasible, it appears as though the answer to the question may nonetheless
be an encouraging yes if an extra nonlinear compression step is 
added to make the analysis doable in practice.
Let us first note that despite the statistical orthonormality property
of a KL-compressed data set, the matrices that need to be inverted to find 
the ML-estimate are in general {\it not} diagonal.
$\B\C\B^t$ is only diagonal at the one point in parameter space where
$\vth=\vth_0$, {\ie}, the point corresponding to the parameter values
that we assumed a priori. The ML-point generally lies elsewhere,
and we need to find it by a numerical search in parameter space, 
so $\B\C\B^t$ will generally not even be sparse. This forces us to 
resort to Cholesky decomposition. For the $n\sim 4000$ case 
of Tegmark \& Bunn (1995), this took about 10 minutes per matrix
on a workstation. Since the time scales as $n^3$ and the 
storage as $n^2$, this would require about 30 years of CPU time 
for $n=10^6$, and about one terabyte of RAM.
Even allowing for the exponential rate at which computer performance
is improving, such a brute force likelihood analysis does not appear 
feasible for a megapixel CMB map within the next ten years.

The crucial question thus becomes how large a compression factor we can
expect to achieve. We will argue that a factor of 10 is just about all
one can do with linear compression, but that quadratic ``pre-compression" 
may be able to gain another factor of 1000 for a next 
generation CMB experiment.

\subsection{The limits of linear compression}

Consider the following simple one-parameter example: we are given 
$n$ numbers $x_i$ drawn from a Gaussian distribution 
with zero mean and an unknown variance $\theta$ that we wish to estimate.
Thus
\beq{TroubleEq} 
\cases{
\vmu	&$=0$,\crr
\C	&$=\theta I$,
}
\eeq
so when we solve \eq{EigenEq}, we find that all the eigenvalues 
are identical: $\lambda_k=1/\theta$. Thus if we compress by keeping only
the first $n'$ modes, the resulting error bar will be 
\beq{TroubleEq2}
\Delta\theta = \sqrt{2\over n'}\theta. 
\eeq
The fact that all eigenvalues are identical means that
we would be quite stupid to throw away any modes at all, 
since they all contain the same amount of information. 
Simple as it is, this example nonetheless illustrates a
difficulty with analyzing future CMB maps.
Even in the best of all worlds, where we had complete sky coverage, we would 
encounter a problem of this type. 
To estimate a multipole $C_\l$, we would be faced with $(2\l+1)$ coefficients 
$a_{\l m}$ with zero mean and unknown variance, just as in the example 
above, which means that the KL-method would be of no use at all
in compressing this data. 

In Bunn (1995) and in one of the above examples, 
it was found that the COBE data set can be 
compressed down to about 
400 numbers without losing much information.
Where does this magic number 400 come from?
In the absence of a galaxy cut, the number would probably be
around 600, since beam smearing has eliminated virtually all information
about multipoles $\l\simgt 25$, and there are about $600$ 
$\alm$-coefficients with $\l<25$. The fact that the galaxy cut removes about 
one third of the sky then reduces the cosmological information
by about a third.
There is also a more direct way to see 
where the compression factor 10 came from.
As mentioned in Section~\ref{LowNoiseSec},
pixelized maps are generally oversampled
by a factor 2.5 or more to avoid aliasing problems, 
which means that the mean pixel separation is
considerable smaller than the beam width. 
Since the sky map is two-dimensional, the number of pixels per
beam area is roughly the square of this number, of 
the order of 10. Future CMB missions will probably 
use similar oversampling rates, which means that a compression 
factor of 10 is probably the most we can hope for with linear 
compression only.

\subsection{Quadratic compression}

Fortunately, we are not limited to linear data compression.
Let us compress the data set of our previous example
into a single number defined by
\beq{QuadraticEq}
y \equiv {1\over n}\sum_{i=1}^n x_i^2.
\eeq
Let us assume that $n$ is very large, say $n\simgt 100$. 
Then $y$ will be very close to Gaussian by the central 
limit theorem. This means that the mean is 
$\expec{y}=n\theta$ and the variance is
$\expec{y^2}-\expec{y}^2=2n\theta^2$. Substituting this into 
\Eq{GaussFisherEq} now gives 
\beq{QuadraticEq2}
\Delta\theta = \sqrt{2\over n+4}\theta \approx \sqrt{2\over n}\theta,
\eeq
{\ie}, the theoretically minimal error 
bar of \eq{TroubleEq2} with $n'=n$. 
(The extra 4 in \eq{QuadraticEq2}, which might seem to indicate
that one can attain smaller error bars than the 
theoretical minimum, should of course be ignored --- it merely
reflects the fact that the Gaussian approximation
breaks down for small $n$.)

In summary, we have found that whereas linear compression
was powerless against our simple toy model, quadratic compression
made mincemeat of the problem, 
condensing all the information into a single number.
This result is hardly surprising, considering that the
compressed data set 
$y$ that we have defined in \eq{QuadraticEq} is in 
fact the ML-estimate of the variance.
It nonetheless has far-reaching implications for the issue of 
how to compress future CMB maps. If we have complete sky coverage, and 
define the compressed data set as 
\beq{QuadraticEq3}
y_\l \equiv {1\over 2\l+1}\sum_{m=-\l}^\l|\alm|^2,
\eeq
then it is easy to show that the new Fisher information matrix 
will be identical
to that of \eq{JungmanEq}, which involved using the entire data set. 
For an experiment with a FWHM beamwidth of 4 arcminutes, there is virtually no information
on multipoles above $\lmax=3000$, so this means 
that in the absence of a galaxy cut, 
we could compress the entire data set into 3000 numbers without losing
any cosmological information at all.
Roughly speaking, this works because the compression throws 
away all the phase information and keeps all spherically averaged amplitude
information, and it is the latter that contains all the information
about the power spectrum. (For non-Gaussian models such as ones
involving topological defects, the power spectrum alone does of course not
contain all the relevant information.)

\subsection{Real-world CMB maps: an outline of a compression recipe}

The discussion above already included the effects of pixel noise and beam 
smearing. Barring systematic errors, there 
are two additional complications that 
will inevitably arise when analyzing future high-quality CMB maps:
\begin{itemize}
\item Foregrounds
\item Incomplete sky coverage
\end{itemize}
Any attempts at foreground subtraction should of course be made before throwing away
phase information, so that available spatial templates of galactic dust, 
radio point sources {\etc} can be utilized. For a recent discussion of 
such subtraction schemes, see {\eg} Tegmark \& Efstathiou (1996). 
The final result of the foreground treatment will almost certainly be
a map where some regions, notably near the galactic plane, have been discarded
altogether. The resulting incomplete sky coverage degrades information
in two different ways:
\begin{itemize}
\item The sample variance increases.
\item The spectral resolution decreases.
\end{itemize}
The former effect causes the variance in individual multipole
estimates to grow as the inverse of the remaining sky area 
(Scott {\etal} 1994),
and this increase in variance is automatically reflected in the final Fisher information matrix.
The latter effect means that the $y_\l$ defined by
\eq{QuadraticEq3}
is no longer a good estimate of
the multipole $C_\l$. Rather, it is easy to show that $\expec{y_\l}$ 
will be a weighted average of all the multipoles $C_\l$.
These weights, known as the {\it window function}, generally form quite a broad
distribution around $\l$, which means that the compressed 
data $y_\l$ are effectively probing a smeared out version 
of the power spectrum.
For a $20^\circ$ galactic cut, this smearing is found to be around  
$\Delta\l/\l \sim 25\%$, which clearly destroys information
about features such as the exact location of the Doppler peaks.

\subsubsection{How to deal with incomplete sky coverage}

Fortunately, the  problem of incomplete sky coverage can be for all 
practical purposes eliminated. 
As described in detail by Tegmark (1996, hereafter T96), it is possible to 
obtain much narrower window functions by simply replacing the 
spherical harmonic coefficients in \eq{QuadraticEq3} by 
expansion coefficients using a different set of functions, obtained
by solving a certain eigenvalue problem.
If is found that for a $20^\circ$ galactic cut, the window functions
widths can be brought down to $\Delta\l\sim 1$ for all $\l$, 
corresponding to a relative smearing of less than a percent at
$\l\sim 200$, the scale of the first CDM Doppler peak.
In other words, as long as none of the models between which we are trying 
to discriminate have any sharp spectral features causing 
the power spectrum
to jump discontinuously from one multipole to 
the next, then virtually
no information at all is lost in our quadratic compression.

In general, smoothing only destroys information if there are 
features on the smoothing scale or below it. 
If the true power spectrum is similar to a CDM spectrum, it will
typically vary on scales $\Delta\l\sim C_\l/(dC_\l/d\l)\sim 100$, at least
for angular scales $\l\simgt 50$.
In other words, even smoothing it with window functions with 
$\Delta\l$ as large as 10 would hardly destroy any information at all
about this part of the power spectrum.
The quadratic compression of T96 produces a spectral
resolution of $\Delta\l\sim 1/\theta$, where $\theta$ is the 
smallest dimension of the patch of sky analyzed, in radians.
In other words, we can allow ourselves even more leeway than 
the galaxy cut dictates without losing information. 
This can be used to save CPU-time in practice.
Since we can achieve the spectral resolution $\Delta\l\sim 10$ with
$6^\circ\times 6^\circ$ patches of sky, we can form compressed
data sets $\y$ separately for a mosaic of such patches covering all 
the available sky, thus radically reducing the size of the matrices that
we need to diagonalize for the T96-method, and then simply
average these different estimates of the power spectrum.

\subsubsection{The final squeeze: KL-method}

Although the above-mentioned prescription will reduce the size of the
data set dramatically, from perhaps $10^7$ numbers to about 3000, 
there will still be considerable amounts of redundancy, since 
power spectrum estimates $y_\l$ for neighboring $\l$-values will be
correlated. Because of this, 
it is worthwhile to subject the new data set $\y$ to a regular
KL-compression. We thus term the above-mentioned
quadratic step {\it pre-compression}: it does by no means need to 
be optimal, and is simply done to reduce the data set
down to a small enough size that it can be fed into the KL-machinery
without practical difficulties.

The values $y_\l$ are Gaussian to an excellent approximation for 
$\l\simgt 50$, by the central limit theorem, 
since they are the sum of $2\l+1\simgt 100$ 
independent numbers.
The remainder, however, are not. Since the
Gaussian property makes such a dramatic simplification both in the
compression step and in the likelihood analysis itself, 
we therefore recommend discarding the $y_\l$-values with $\l<50$ and
replacing them by the 2500 spherical harmonic coefficients 
$\alm$ with $\l<50$ before proceeding to the KL-compression step.
This way, the entire data set $\y$ will be Gaussian. In addition, 
no information whatsoever has been lost about the largest angular
scales, where some models in fact do predict rather sharp 
spectral features which could render the quadratic compression step
inappropriate.

In summary, we argue that the following prescription for 
analyzing future $10^7$ pixel CMB map will be fairly close to optimal:
\begin{enumerate}
\item Foregrounds are subtracted making maximal use of multifrequency data
and spatial templates obtained from other experiments.
\item The most contaminated regions, for instance the galactic plane, are 
discarded altogether.
\item The remaining sky is expanded in spherical 
harmonics up to $\l=50$ and the coefficients saved.
\item This remaining sky is covered by a mosaic of overlapping regions, 
whose diameter are 
of order $5^\circ-10^\circ$.
\item The angular power spectrum is estimated separately from each of 
these patches with the method
of T96, up to $\l=3000$.
\item All these power spectrum estimates are averaged.
\item The first 50 multipole estimates are discarded, and replaced by the
2500 numbers from step (iii).
\item The resulting data set (about 5500 numbers) is compressed 
with the KL-method.
\item All cosmological parameters of interest are estimated jointly from this 
compressed
data set with a likelihood analysis.
\end{enumerate}

\subsubsection{Open problems}

The above prescription for quadratic compression was necessarily
rather sketchy, since a detailed treatment of these issues 
would go well
beyond the scope of this paper. Indeed, the discussion above left
several important questions unanswered, which are clearly 
worthwhile topics for further research. 
Here are two such examples:

\begin{itemize}

\item How can information loss during pre-compression be minimized?
Above we merely showed pre-compression to be lossless in the absence
of noise (or with identical noise levels in all pixels). 
In the presence of noise, one would expect lossless compression 
to involve some form of inverse-variance pixel weighting, 
\ie, giving less weight to noisier pixels. On the other
hand, pushing such noise weighting too far could broaden the window 
functions to the point where low spectral resolution led to 
irreversible information loss. 

\item
How is quadratic precompression best implemented numerically?
For instance, are there particularly choices of shapes and sizes of
the above-mentioned patches that substantially facilitate
the calculation of the final mean vector and correlation matrix?
Is a direct analytic calculation of these quantities feasible
($\C$ involves terms quadratic in the power spectrum), or is it 
faster to resort to Monte Carlo techniques for this step?

\end{itemize}

\subsection{Real-world galaxy surveys}

Also for large future galaxy surveys, some form of
precompression appears to be necessary before 
a KL-compression can be done.
If we wish to retain all the information on clustering down to say 
$3h^{-1}\Mpc$ scales in a 3D volume of typical dimension $300 h^{-1}\Mpc$, 
we clearly need to ``pixelize" it into at least $(300/3)^3 = 10^6$ numbers.
Since even the power spectrum in the deeply non-linear regime contains
valuable cosmological information, it would not seem justified
to simply ignore these small scales.

Unfortunately, the galaxy survey problem is considerably more difficult 
than the corresponding CMB problem in that information about, 
for instance, redshift space distortions 
lies hidden not merely in the overall power spectrum, but also in the
phases, in the differences between radial and transverse clustering patterns.


As we have noted, transforming to weighted harmonic amplitudes 
is one way of reducing the 
number of data ``points'' without
losing cosmological information
For instance, in the
case of the analysis of the 1.2Jy survey by HT95 and BHT95, only
the first 1208 modes where used from the 2000 galaxies,
with the limit on modes used being set by the survey size 
and the smallest scale still in the linear regime.
However, if the full range of available modes is desired,
we might need higher order compression options.

One way to implement quadratic compression, while 
retaining the important phase information, is to transform
to some coordinate basis which is orthogonal in redshift--space.
We can then apply the quadratic compression to estimate a
``power spectrum'' in the transformed space, having lost none 
of the underlying phase information. Some progress along
the lines of finding an orthonormal basis function for 
redshift space has been made by Hamilton \& Culhane (1995).
However, these methods still fail to adequately deal with 
shot--noise and the phase mixing introduced by a finite 
angular mask function.

Thus the issue of whether one can do even better with galaxy surveys, 
while staying within the realms of numerical feasibility, 
remains a challenging open question.

\section{CONCLUSIONS}
\label{ConclusionsSec}

We have given a comprehensive discussion of how to best estimate a set of 
cosmological parameters from a large data set, and concluded the following:
\begin{itemize}

\item Well-known information-theoretical results roughly speaking state that
a brute-force likelihood analysis using the entire data set gives
the most accurate parameter determination possible.

\item For computational reasons, this will be unfeasible for the next 
generation of CMB maps and galaxy surveys.

\item The solution is to use a good data compression scheme.

\item The optimal {\it linear} compression method is the 
Karhunen-Lo\`eve method, of which the so-called signal-to-noise eigenmode
method is a special case.

\item Although the standard KL-method applies only when 
 estimating a single parameter,
it can be generalized to the multi-parameter case by 
simply adding a step consisting of a singular value decomposition (SVD).

\item This SVD step also provides a simple way out of the 
Catch-22 situation that
one needs to specify the parameter values before one has measured them.

\item The KL-method produces a compression factor $\sim 10$ for 
typically sampled 
CMB maps, and also for the redshift space distortion analysis
of Heavens \& Taylor (1995).

\item However, this is not enough to handle a high-resolution 
next generation CMB map.

\item Fortunately, it appears as though this can be remedied 
by adding a quadratic pre-compression step without substantial 
information loss.

\end{itemize}
Cosmology, which used to be a data-starved science, is now 
experiencing a formidable explosion
of data in the form of both CMB maps and galaxy redshift surveys.
Around the turn of the millennium, we will probably be equipped with data sets so rich
in information that 
most of the classical cosmological parameters can 
--- in principle --- be determined with accuracies of 
a few percent or better.
Whether this accuracy will be attainable also in practice depends
crucially on what data-analysis methods are available.
We have argued that the prospects of achieving this accuracy 
goal are quite promising, 
especially on the CMB side (which is slightly simpler), by using
a multi-parameter generalization of the Karhunen-Lo\`eve method 
together with a quadratic pre-compression scheme.
However, much work remains to be done on precompression issues
to ensure that we can take full advantage of the wealth of data
that awaits us.

\bigskip
\noindent{\bf ACKNOWLEDGMENTS}
\smallskip

\noindent
The authors wish to thank Ted Bunn,
Andrew Hamilton and our referee, Uros Seljak,
for useful comments and suggestions.
This work was supported by
a PPARC research assistantship (to ANT), 
European Union contract
CHRX-CT93-0120, Deutsche Forschungsgemeinschaft
grant SFB-375 and 
NASA through a Hubble Fellowship,
{\#}HF-01084.01-96A, awarded by the Space Telescope Science
Institute, which is operated by AURA, Inc. under NASA
contract NAS5-26555.

\bigskip
\noindent{\bf REFERENCES}

\bib Bennett, C. L. {\etal} 1996, ApJ, 464, L1

\bib Bond, J. R. {\etal} 1994, Phys. Rev. Lett., 72, 13

\bib Bond, J. R., 1995, Phys. Rev. Lett., 74, 4369

\bib Bunn, E. F. 1995, Ph.D. Thesis, U.C. Berkeley, 
{\it ftp pac2.berkeley.edu/pub/bunn}

\bib Bunn, E. F., Sugiyama N. 1995, ApJ, 446, 49

\bib Bunn, E. F., Scott, D. \& White, M. 1995, ApJ, 441, L9
     
\bib Davis M., Peebles P. J. E., 1983, APJ, 267, 465

\bib Dodelson, S., Gates, E. \& Stebbins, A. 1996, ApJ, 467, 10 

\bib Feldman H. A., Kaiser N., Peacock J. A., 1994, APJ, 426, 23

\bib Fisher, R. A. 1935, J. Roy. Stat. Soc., 98, 39

\bib Fisher K. B., Huchra J. P., Strauss M. A., 
Davis M., Yahil A., Schlegel 
D., 1995, ApJS, 100, 69

\bib G\'orski, K. M. 1994, ApJ, 430, L85

\bib Gunn, J., Weinberg, D., 1995,
in {\it Wide-field spectroscopy and the distant universe}, proc. 35th 
Herstmonceux conference,
eds S.J. Maddox \& A. Arag\'on-Salamanca, World Scientific, p3

\bib Hamilton, A. J. S, Culhane, M., MNRAS, 278, 73

\bib Heavens, A. F., Taylor, A. N., 1995, MNRAS, 483, 497

\bib Hu, W. \& Bunn, E. F. \& Sugiyama, N. 1995, ApJ, 447, L59
     
\bib Hu, W., Scott, D., Sugiyama, N. \& White, M. 1995, 
Phys. Rev. D, 52, 5498
 
\bib Jungman, G., Kamionkowski, M., Kosowsky, A. \& Spergel, D. N. 1996a,
Phys. Rev. Lett., 76, 1007

\bib Jungman, G., Kamionkowski, M., Kosowsky A. \& Spergel D. N. 1996b,
Phys. Rev. D, 54, 1332

\bib Kaiser, N. 1987, MNRAS, 227, 1

\bib Karhunen, K., {\it \"Uber lineare Methoden in der 
Wahrscheinlichkeitsrechnung} (Kirjapaino oy. sana, Helsinki, 1947).

\bib Kendall M. G., Stuart, A., 1969.  The Advanced Theory of Statistics, 
Volume II, Griffin, London

\bib Kenney, J. F. \& Keeping, E. S. 1951, {\it Mathematics of Statistics, Part II},
2nd ed. (Van Nostrand, New York).

\bib Knox, L. 1995, Phys. Rev. D, 52, 4307

\bib Lawrence, A., Rowan-Robinson, M., Saunders, W., 
Parry, I., Xia Xiaoyang, 
Ellis R. S., Frenk, C.S., Efstathiou, G. P., Kaiser, N., Crawford, J., 1996, 
to be submitted to MNRAS

\bib Liddle, A. R. {\etal} 1996, MNRAS, 281, 531 

\bib Peacock, J. A. \& Dodds 1994, MNRAS, 267, 1020

\bib Press, W. H.,  Teukolsky, S. A., 
Vetterling, W. T., Flannery, B. P., 1992, 
Numerical Recipes. Cambridge University Press, Cambridge

\bib Saunders, W. {\etal} 1996, in preparation
   
\bib Scott, D.,  Srednicki, M. \& White, M. 1994, ApJ, 421, L5

\bib Seljak, U. \& Bertschinger, E. 1993, ApJ, 417, L9

\bib Seljak, U. \& Zaldarriaga, M. 1996, preprint astro-ph/9603033

\bib Smoot, G. F. {\etal} 1992, ApJ, 396, L1

\bib Taylor, K., 1995,
in {\it Wide-field spectroscopy and the distant universe}, proc. 35th 
Herstmonceux conference, eds S.J. Maddox \& A. Arag\'on-Salamanca, 
World Scientific, p15

\bib Tegmark, M., Bunn, E. F. 1995, ApJ, 455, 1

\bib Tegmark, M. 1995, ApJ, 455, 429

\bib Tegmark, M. 1996, MNRAS, 280, 299
     
\bib Tegmark, M. \& Efstathiou, G. 1996, MNRAS, 281, 1297		

\bib Therrien, C. W. 1992, {\it Discrete Random Signals and Statistical Signal Processing}
(Englewood Cliffs: Prentice Hall)

\bib Vogeley, M. S. 1995, in ``Wide-Field Spectroscopy and the Distant Universe",
eds. Maddox \& Arag\'on-Salamanca (World Scientific, Singapore)

\bib Vogeley, M. S. \& Szalay, A. S., 1996, ApJ, 465, 34 

\bib White, M. \& Bunn, E. F. 1995, ApJ, 450, 477

\bib Yamamoto, K. \& Bunn, E. F. 1996, ApJ, 464, 8 

\end{document}